# Influence of refractive index in light scattering measurements of biological particles

Sanchita Roy[1,*], Jamil Hussain[1], Semima Sultana Khanam[1], Showhil Noorani[1] and Aranya B. Bhattacherjee[1,2]

[1]*Department of Physics, School of Applied Sciences, University of Science and Technology, Meghalaya, District- Ri Bhoi, 793101, Meghalaya, India.*
[2]*Department of Physics, Birla Institute of Technology and Science, Pilani, Hyderabad Campus, Telangana 500078, India*

[*]*e-mail:* rsanchita1@gmail.com

## Abstract

Investigations on diverse particulate matter can be carried out by non-invasive and non-destructive analytical means of light scattering technique. Exploration of small particles using this tool must address the concept of refractive index accurately. Interpretation of light scattering results for quantification of morphology of biological particles like viruses and bacterial cells can only be validated by exact evidence of its refractive index. Size and shape quantification of such particles by collecting information from the scattered light, was reported earlier by many researchers but parameter like refractive index was considered as a standard value employed from literature or selected as a median value of biological particles which are alike. A strong analysis is needed on inclusion of appropriate value of refractive index for revelation of light scattering results. This important perception is otherwise not considered strongly in similar works reported earlier. We primarily took an attempt to find the influence of refractive index in light scattering studies with allusion to biological particles like *Staphylococcus aureus*, *Escherichia coli*, *Corona virus* (*SARS COV 2*) etc.

## 1. Introduction

The determination of refractive indices is important in wide range of fields. Most notably, for size, shape, etc. of the biological particles using light scattering, an estimate of refractive index (RI) must be given to calculate the particle size from the scattering data [1-5]. In the biological realm, refractive index is important in terms of phase contrast microscopy, flow cytometry, and light scatter-based detection methods [6]. The non-invasive and non-destructive handling allowed by light scattering mechanism can be relevant to an extensive range of species of biological particles [7,8].

The refractive index '$n$' of a material is a parameter that well describes the unique interaction between the material and an electromagnetic wave [18, 19]. Knowledge of the interaction of light with matter allows us to predict the propagation of light to harness the properties of light for applications such as photonics, imaging, information processing, etc. [9, 10]. More so, knowledge of the refractive indices of two different materials provides a way of predicting the scattering of light at the interface of the materials. For particles having high real indices, the size quantification depends less on the imaginary component of the refractive index. Again, as size of the particle increases to nearly above 10 micrometers to 20 micrometers, Fraunhofer Diffraction is suitable to describe particle size and subsequently the refractive index becomes important to a lesser extent [11, 12].

The penetration depth of an electromagnetic field is a function of wavelength and the attenuation constant for the field is proportional to the imaginary part of the refractive index of the material. The following relationship of refractive index holds true based on the above information:

i.e., RI = $n + i\,k$, where:

$n$ = the real component of refractive index

$c$ = speed of light in vacuum

$k$ = the extinction coefficient of the medium or material

Here, the refractive index is defined by two components real and imaginary, where the real part '$n$' refers to the phase velocity whereas the imaginary part '$k$' refers to the extinction coefficient and also mass attenuation coefficient sometimes [13,14]. The imaginary component of the refractive index parameter is thus defined as the reduction of transmission of optical radiation, i.e. $k = (\lambda/4)\eta$, where:

$\eta$ = the absorption coefficient

λ = the wavelength of light

The absorption coefficient (η) is the reciprocal of the distance light will penetrate the surface of the material and gets attenuated to 1/e (about 37%) of its original intensity at the surface [11]. Thus, it must be noted that refractive index is a function of frequency. Its imaginary part is often not mentioned or not taken into account because it is essentially considered to be zero for transparent dielectrics [5, 11,12]. Biological cells can be contemplated as dielectric objects which has refractive index distribution [18]. Simulations of light scattering delivers an efficient tool for perusing nature of scattering as well as the morphology of cells. The scattered light signals carry vital information about characterization methods of the biological particles, which otherwise becomes cumbersome using biochemical methods. It is important to note that the cell refractive index (RI) is an intrinsic optical parameter which controls light propagation through the cell matrix. The RI of cell is considerately associated with distribution of its mass. Thus, it has the ability to provide significant comprehensions for varied biological models. The different techniques of measurement established earlier could only gauge the effective RI of a cell or cell suspension, thus only inadequate substantiation of the refractive index of cell could be attained [15]. This is a limitation for its detailed exploration and connection with the actual findings [ 15].

There are many theoretical approaches for solving light scattering problems. One important theory is the Mie theory which describes light scattering by spheres [ 2,4]. Mie theory is also applicable to randomly oriented non-spherical particles. How the imaginary part of refractive index of spheres affects the way it scatters light was narrated by other researchers [16, 17]. It was described that for spheres of arbitrary radius $R$ and complex refractive index ($n+ik$), how a new parameter indicated the imaginary part '$k$' which severely affects the scattering by a sphere by using $Q$-space analysis [16, 17]. Thus, inclusion of imaginary part decides the scattering by spheres, however, it requires detailed analysis whether it is valid for bio-particles which are spherical in nature. While other works reported reflectance spectrometry which was in association with the Kramers-Kronig analysis used to determine the complex refractive index of biological cells [20]. With this procedure, the real and imaginary parts of RI can be simultaneously determined. In other words, this combination is a convenient approach to substantiate the complex RI of biological cells during the cell cycle [20]. This points out the necessity of inclusion of unignorable change of the imaginary part of index of refraction associated with each phase of the cell for light scattering investigations. Thus, the scattering from different stages of cell ( in cell cycle) of the same strain might possibly display significant unique scattering profile[20]. This indicates that even if we ignore the imaginary part of index of refraction during the light scattering investigations of biological particles, it would only provide inadequate information of the scatterers. Hence, the characterization methods may not provide real predictions. In other statements, it has been reported that light scattering investigations carried out considering same median value of refractive indices ($n =1.06$, for visible range) of different strains of individual biological cells can be used for modelling or characterization of such bio-particles [ 21, 22]. But this cannot justify the uniqueness of scattering profile of each type of cell even after considering that there exist only elusive biochemical and structural differences between bacterial species. It is very hard to find information about the refractive index and its spectral change for viruses as reported by other workers [21,22]. This indicates the feasibility to study the effects of only shape of particles by using the median value of RI parameter [21, 22]. This fact reveals the need of exact information about the complex RI for any bacteria or viruses for their morphological or optical characterization using light scattering tool.

## 2. Methodology

The scattered light intensity by a particle, or a volume of particles represents a function of the wavelength of incident light and angle between incident and scattered radiation. Light scattering techniques accurately determined structures within biological samples and are increasingly developed as biomedical diagnostic tools. Biomedical application of light scattering provides researches and studies in the field of Bio-photonics with engrossed information on light scattering for biomedical applications [23-25, 10].

In this work, we present a review of implications of refractive index with reference to biological cells in light scattering studies. In utmost cases, biochemical process is used to characterize and analyze such particles. But most often these methods either changes its morphology or destroys the sample. This reason makes light scattering an important investigative tool because it is a non-destructive method to characterize and analyze such particles [5,7,8]. Furthermore, it has been reported that the morphology of bacteria also affects their refractive index [26]. It can have two unlike settings with diverse refractive indices, thereby portraying clear variation between their vegetative and sporulation forms [26]. It has also been reported that since the viral particle is extremely minute in size, it is tougher to measure its RI [26]. The measurement procedures of its RI are limited and primarily based on suspension measurement. Due to the restriction of the measurement techniques, adequate information on literature cannot be found for refractive index measurement of viruses [26]. Also there has been significant report on the map of distribution of refractive index as a function of the 'Z' coordinate at different pixels at certain horizontal line for *Coronavirus (SARS-CoV-2)* (the same SEM image was considered for our investigation also as shown in figure 1(c)) [27]. This provides a new general information that there may be deviation of the actual refractive index from the reported one. Thus, demands inclusion of accurate refractive index for generating the scattering profiles (both real and imaginary parts), which can quantify the size and shape of the particles.

We considered biological cells like *Staphylococcus aureus* or *S.aureus*, *Eshcherichia coli* or *E. coli* and *Corona virus (SARS COV2)* for our study. *S.aureus* is a round-shaped Gram-positive bacterium . It has been reported that *S. aureus* which enters a varied kind of host cell kinds is capable of duplicating and persisting within those host cells [28]. It is a human pathogen which has caused many epidemic outbreaks over the past 100 years and can turn into a dangerous pathogen [7]. *E. coli* is a short rod-shaped bacterium whose length and diameter ranges from 50 nm to 250 nm and 5nm to 25 nm respectively. It is the most common pathogen and a common cause of infections in human which can be life- threatening [24, 29]. Corona viruses (SARS-CoV-2) are group of RNA viruses. These bio-particles are round-shaped bearing protein spikes. It is a hostile pathogen which eroded a mass human population around the globe. WHO declared global pandemic in 2019 and 2020 due to its alarming outbreak. It needs a proper diagnostic tool to arrest pathogenic spread of such strains because it concerns human life and economic loss [22, 31]. The study of characteristic properties of such strains by light scattering can provide new tools for its rapport dealing with various issues concerning global health.

In this work we also took an attempt for comparative analysis of light scattering profiles considering refractive index parameter of some selected biological cells like *E.coli, S. aureus and Corona virus ( SARS COV-2)*. RI is an important parameter in light scattering intensity profile generation. The analyses included selecting a biological cell and finding out the signatures of light scattering choosing same modal size but different refractive and also for different modal radius but same refractive index. Mie scattering plots were obtained by means of the standard light scattering software developed by Philip Laven's Mieplot v4.304 [32]. Refractive index of these particles was found from different sources in literature [7, 12, 21, 22, 30].

The volume scattering function (VSF) given by β(θ) describe the angular distribution of the scattered light intensity [1- 4]. For an ensemble of particles illuminated by an unpolarised light, the VSF is obtained by integrating $S_{11}$ element of the Mueller matrix over the size distribution *N(a)* [1- 4,7,8]. The volume scattering function of these biological particles were obtained theoretically [32]. Average differences are expected in their dielectric composition when bacterial cells compared with viruses.

## 3. Scanning Electron Micrograph (SEM) image of bio-particles

The SEM image of *E. coli, S. aureus* and *Coronavirus (SARS-CoV-2)* were collected from different source [ 7,8, 24, 22]. Modal radii of each sample were calculated from the SEM images and the parameter was incorporated to generate the scattering profile using Mie plot. Figures 1(a), 1(b) and 1 (c) exhibits the SEM images of *E. coli, S. aureus* and *Coronavirus (SARS-CoV-2)* respectively [7]. The refractive indices of these particles are tabulated in Table 1.

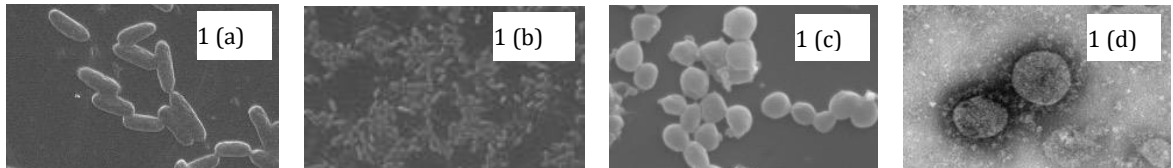

Figure 1: (a) & (b) SEM image of *E.coli at magnification of X10,000 and X3500 respectively* (c) *S. aureus at magnification of X20,000* (d) *Coronavirus (SARS-CoV-2)* Photo courtesy of CDC/ Fred Murphy)

| | Table 1: Refractive index parameter corresponding to the bio-particles | | | | | |
|---|---|---|---|---|---|---|
| Name of the Bio-particle | Shape | Orientation | Aspect ratio | Refractive index $'n'$ | $'\delta n'$ | Wavelength $'\lambda'$ |
| *Eshcherichia coli* | Rod-shaped | Random | 0.42 | 1.38 | - | 632.8 nm |
| *Staphylococcus aureus* | Spheroidal | Random | 0.58 | 1.10 | 0.28 | 632.8 nm |
| *Corona virus(SARS-CoV-2)* | Spherical | Random | 0.77 | 1.06 | 0.04 | 632.8 nm |

We have generated Mie plots for *S. aureus, E. coli and Coronavirus (SARS-CoV-2)* bio-particles at the wavelength of $\lambda = 632.8\ nm$ was chosen because this wavelength predominate the non-resonant absorption [7,24,27, 34]. Single scattering is considered for our studies based on the distribution of particles observed from the SEM images (figures 1(a) to 1(d). By choosing the important parameters from SEM images and from literature review, scattering profiles were generated. Figures 2(a) exhibits the comparative profiles of *E. coli*, *S. aureus* and *Coronavirus (SARS-CoV-2) at* $\lambda = 632.8\ nm$ **.** Figure 2(b) exhibits the Intensity vs scattering angle profiles of *E. coli* for different modal radii but same value of RI. Figure 2 (c) exhibits the profiles of. *S. aureus* at different RI but same modal radii. Figure 2(d) represents profiles for *E. Coli* at different RI but same modal radii and Figure 2 (e) represents *Corona virus (SARS COV-2)* at different RI but same modal radii**.** The information of modal radii was obtained from the SEM images and the values of RI was taken from sources reported earlier (for specific biological cells) [ 7, 21, 23, 24]. All the scattering profiles (Figures 2(a) to 2(e)) were normalized. From figure 2(a) we see that the scattering profiles of the bacterial cells follow similar trend line, however Mie fluctuations were present. This may be attributed to the modal radius being incorporated into the program irrespective of the narrow size distribution observed in both the samples. Minimal backscattering was more pronounced for the *S. aureus* cells. The scattering profile of the *Corona virus* deviated to a greater extent from that of the other two profiles. The characteristic profile of the virus is more uniform over the total angular range. It must be noted that the actual characteristics can be validated only by comparative analyses with experimental results because during actual scattering from sample, the net scattering will be the volume scattering from the size distributed particles. Nevertheless, while generating the theoretical profile, only one size parameter (in this case, the modal radius) is incorporated into the program. From figure 2(b), we see that the scattering profiles of *E. coli* for different modal radii (corresponding to the figures 1(a) & 1(b)) and same value of *n* = 1.38 exhibits unique scattering profile although they follow a similar trendline. Mie fluctuations were more prominent for particles having greater value of modal radius in comparison to the other profile. It is noteworthy to mention that even if the values of RI were same but choice of the modal radius contributes to the deviation and distinctness of the profiles. In figure 2(c), the intensity vs scattering angle of *S. aureus* at different values of '*n*' i.e. 1.1 and 1.4 respectively [ 7, 12]; same modal radii= 150 nm at λ= 632.8 nm is presented. Although, $\delta n$ = 0.3 which is very insignificant, but the scattering profiles vary from each other. Both the profiles exhibit backscattering. The uniqueness of scattering profiles corresponding to the different values of RI emphasizes on importance of exact information of RI when it concerns biological particles. Slight discrepancies also might play a vital role in tangible morphological characterization. In Figure 2 (d), the intensity vs scattering angle of *E. Coli* at different 'n', i.e. 1.38, 1.39 & 1.40 were considered, $\delta n = 0.01$, having same modal radii = 440 nm at λ= 632.8 nm. The three different values of RI were chosen based on the information that usually RI of bacterial cells ranges from 1.38 to 1.40 [7,8,12, 21,24]. It has been observed that there is a great correspondence between all the three scattering profiles with only small marginal difference.

Figure 2(e) depicts the variation of β(θ) as a function of scattering angle θ for *Corona virus (SARS COV-2)* with three different refractive index '*n*' and having fixed value of modal radius. As evident from the figure, the plots of variation of β(θ) for the three values of '*n*' approximately coincides. In order to understand this peculiar behavior, we refer to figure 2 (a) where we clearly observe that variation of β(θ) as a function of scattering angle θ for *Corona virus (SARS COV-2)* does not deviate significantly from β(θ) ~ 1.0 (a.u.) (including closer values to the normalized angle) for fixed '*n*' and modal radius as compared to that for *E.coli* and *S. aureus*. This is possibly because of the relatively smaller dimension (modal radius =120 nm) of *Corona virus (SARS COV-2)*. The relatively small dimension of the *Corona virus (SARS COV-2) does* not practically satisfy the criterion of Mie theory entirely. Here, $\delta n = 0.03$ & 0.65 relative to $n = 1.06$. The choice of $n = 1.06$ is based on the fact that RI of most of the viruses are taken as a median value [ 20-23, 27]. Although $\delta n$ differs but the scattering profiles shows greater correspondence between them. It is also important to note that in all the above-mentioned theoretical and experimental studies, the imaginary part of RI was not taken into account. Thus, it indicated that if the extinction coefficient is included then there is a possibility that the plots of variation of β(θ) for different '*n*' may not exhibit closeness as it has been observed in our present results.

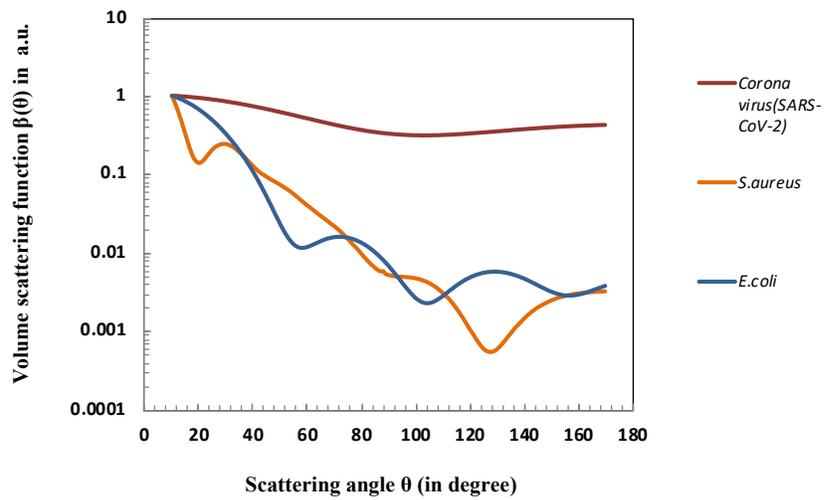

Figure 2(a) : Intensity vs scattering angle of *S. aureus, E.coli* and *Corona virus (SARS COV2)* at λ= 632.8 nm

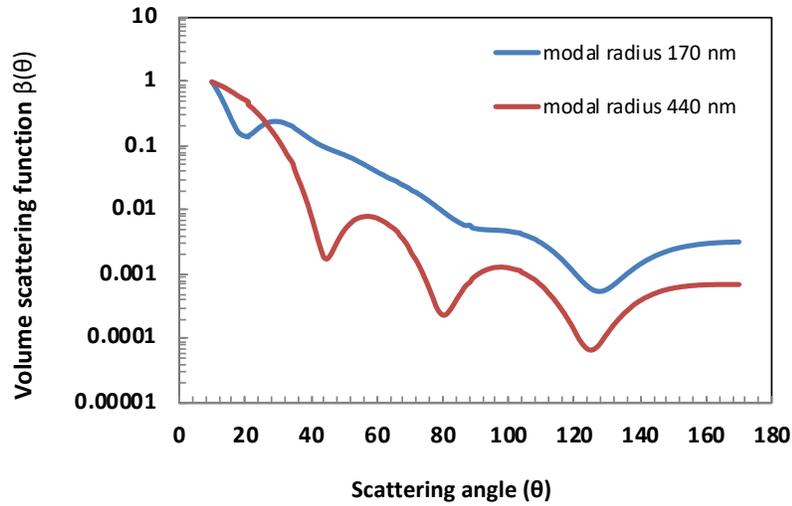

Figure 2 (b): Intensity vs scattering angle of *E. coli* for different modal radii; same value of '$n = 1.38$' at λ = 632.8 nm

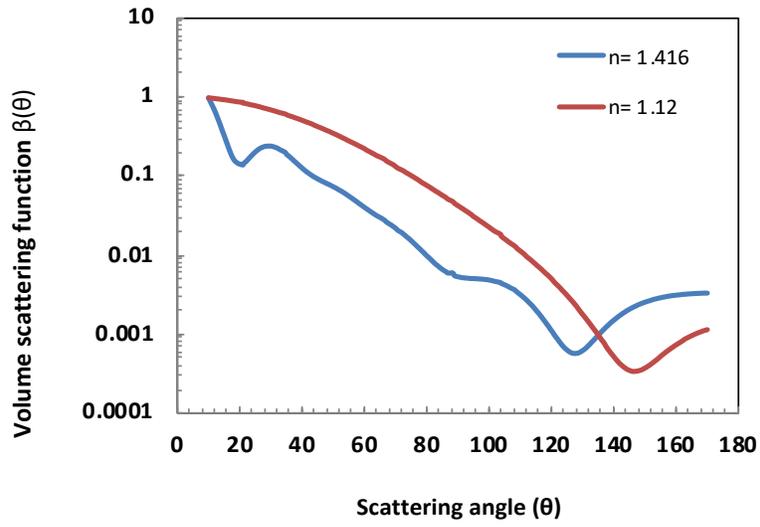

Figure 2(c): Intensity vs scattering angle of *S. aureus* at different '*n*'; same modal radii= 150 nm at λ= 632.8 nm

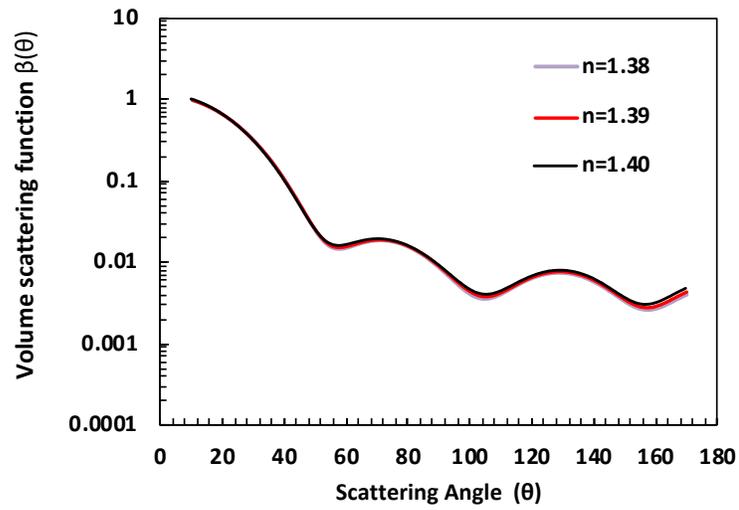

Figure 2 (d): Intensity vs scattering angle of *E. Coli* at different '*n*'; same modal radii= 440 nm at λ= 632.8 nm

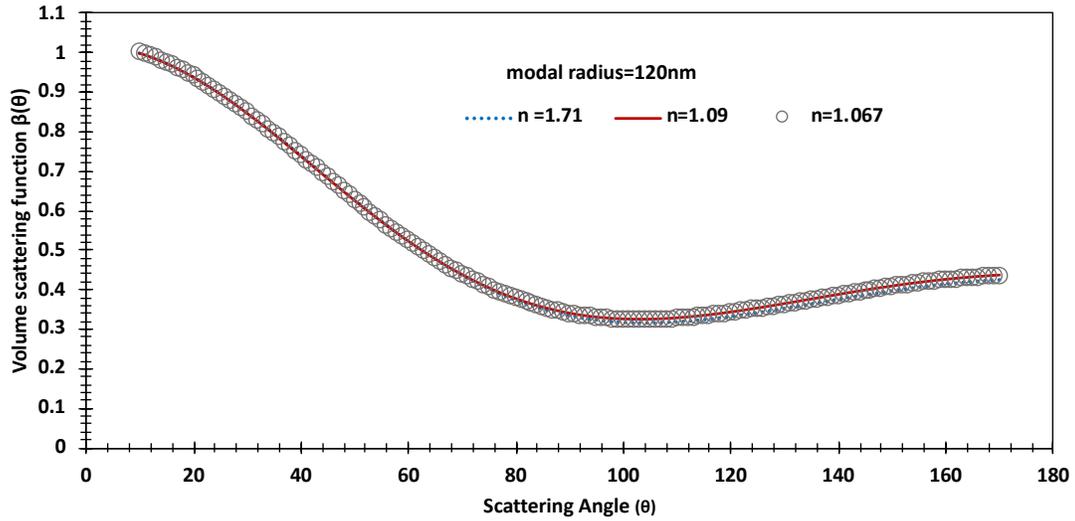

Figure 2 (e): Intensity vs scattering angle of *Corona virus (SARS COV-2 )* at different 'n'; same modal radii = 120 nm at λ= 632.8 nm

## 4. Results and Discussions

The results of the scattered light intensity of *S. aureus*, *E.coli* and *Corona virus (SARS COV-2)* are shown in figures 2(a) to 2(e). Figure 2(a) shows unique signatures of these cells even with small differences in $\delta n$. It can be observed from the figures that the scattering pattern portray number of perceptible features depending on the RI chosen. Experimental light scattering profile of *S. aureus* exhibited similarity in the scattering pattern as compared to the theoretical scattering profile, however minor differences cannot be overlooked. Our constraint is on experimental scattering results of *E.coli* and *Corona virus (SARS COV-2)*. Comparative analyses with the experimental scattering profile of *E.coli* and *Corona virus (SARS COV-2)* will provide a better insight to our observations and results. However, it is indicated by our observations that refractive index parameter which was selected as a median value for all similar biological cells or standard value implemented from other reports may not provide complete or actual information of the scatterers. Scattering profiles does depend upon number of features like the modal radius corresponding to RI of the bio-particle, dynamism of RI corresponding to cell cycle, scattering profile corresponding to range of RI distribution based on pixel of SEM image, consideration of size and shape distribution rather than single size parameter in the software program etc.

Contemplation of median value of refractive index in similar bio- particles (viruses) or consideration of RI value from a range of refractive index ($\delta n$ ranging from 0.01 to 0.65 for different bio-particles considered) is a limitation in actual morphological characterization. Comparison with experimental results deliberating the same parameters can further validate our observations because average differences in scattering profiles are expected due to their dielectric composition when bacterial cells compared with viruses. Although it has been reported by other researchers how size and shape parameters of bio-particles could be quantified by light scattering tool but it must be noted that inclusion of effects of imaginary part of RI (negligible) and the other factors discussed above may provide subtle differences but definiteness of the scattering profiles of these bio-particle.

## 5. Conclusion

We attained a general inference that merely by considering a relative refractive index or median refractive index or by adopting any value of RI from a range of RI concerning biological cells may not provide the actual characteristic scattering profile. The value of $\delta n$ significantly influences the scattering profile. The smaller is the value of $\delta n$( same modal radius), greater is the correspondence in scattering profiles, and when greater is the value of $\delta n$ the differences become more pronounced in the signatures. When $\delta n = 0$, we expect certainty in correspondence of scattering profiles for same biological cells (with same parameters discussed), but the fact is that we may develop distinctive unique characteristic profile with different modal radius. Thus, we state that proper choice of refractive index in light scattering measurements (theoretical, experimental or modeling) with allusion to biological particles is very important factor for characterizations. Considerations of these points directs towards the possibility of recording dynamic light scattering in addition to static light scattering from biological cells for further validation of our results.

**Acknowledgement**

The author wish to thank the Research division of University of Science and Technology Meghalaya for support.